\documentclass[final,3p,times,twocolumn,sort&compress]{elsarticle}  
\usepackage{graphicx}  
\usepackage{dcolumn}   
\usepackage{bm}        

\usepackage{amsmath}
\usepackage{amsthm}
\usepackage{amssymb}
\usepackage{amsfonts}

\usepackage{color}
\definecolor{orange}{RGB}{255,127,0}

\usepackage{bbm}
\usepackage[format=plain,justification=raggedright,textfont=small]{caption}

\begin{document}


\title{On the effect of time-dependent inhomogeneous magnetic fields \\ in
electron-positron pair production}
                             
\author[jena]{Christian Kohlf\"urst}
\ead{christian.kohlfuerst@uni-jena.de}
\author[graz]{Reinhard Alkofer}
\ead{reinhard.alkofer@uni-graz.at}                             
        
\address[jena]{Theoretisch-Physikalisches Institut, Abbe Center of Photonics, \\
Friedrich-Schiller-Universit\"at Jena, Max-Wien-Platz 1, D-07743 Jena, Germany \\ 
Helmholtz-Institut Jena, Fr\"obelstieg 3, D-07743 Jena, Germany}        
\address[graz]{Institute of Physics, NAWI Graz, University of Graz, A-8010 Graz, Austria}          
        
\date{\today}

\begin{abstract}
Electron-positron pair production in space- and time-dependent electromagnetic fields is
investigated. Especially, the influence of a time-dependent, inhomogeneous magnetic field 
on the particle momenta and the total particle yield is analyzed for the first time. 
The role of the Lorentz invariant {$\mathbf{E}^2 - \mathbf{B}^2$}, including its sign and local
values, in the pair creation process is emphasized. 
\end{abstract}


\maketitle

\paragraph{Introduction}

Although already predicted in the first half of the last century \cite{Sauter:1931zz}
electron-positron pair production attracted renewed attention over the last decade.
This interest is strengthened by experiments verifying the possibility of
creating matter by light-light scattering \cite{Burke:1997ew}.
Upcoming laser facilities, {\it e.g.}, ELI \cite{eli,Heinzl:2008an} and
XFEL \cite{xfel,Ringwald:2001ib}, as well as newly proposed
experiments \cite{Marklund:2008gj} are expected to 
deepen our understanding of matter creation from fields.

Note that in the very special case of constant and homogeneous
fields the Lorentz invariants 
\begin{equation}
 \mathcal F = \frac{1}{2} \left( \mathbf{E}^2 - \mathbf{B}^2 \right) ,\quad \mathcal G = \mathbf{E} \cdot \mathbf{B}
\end{equation}
determine the particle production rate \cite{Dunne:2004nc}. In {constant crossed fields} 
$\mathcal G$ vanishes which highlights then the role of 
the action density $\mathcal F$ in pair production.

Although electric and magnetic fields appear in equal magnitude in the quantity $\mathcal F$
magnetic fields are usually ignored in theoretical investigations of pair production.
This may be {motivated} by the fact that for perfect settings the magnetic field 
vanishes in the overlapping region of two colliding laser beams \cite{Alkofer:2001ik}. 
Hence, the majority of studies on pair production have examined this process for
time-dependent electric fields only \cite{PhysRevLett.101.130404,abc}. (NB:
Configurations with an additional constant magnetic field have been investigated in \cite{Dunne:1997kw}.)

But in studies of pair production by electric fields it turns out that exactly the time-dependence of the fields is most influential, and depending on it one observes different 
mechanisms behind pair production \cite{Nousch:2012xe,Diss}. 
In a first, almost  superficial, way one can distinguish multi-photon pair production 
\cite{Kohlfurst:2013ura,Ruf:2009zz} from the Schwinger effect 
\cite{Hebenstreit:2011pm,Cohen:2008wz}. {Employing} multi-timescale fields,
{however}, a rich phenomenology opens up. Hereby, {\it e.g.}, the dynamically-assisted Schwinger effect \cite{PhysRevLett.101.130404,Orthaber:2011cm,Linder:2015vta,Panferov:2015yda}
is only one, although the most prominent, example.

Given this situation, and in view of realistic possibilities of an experimental 
verification, it is an unsatisfactory situation that so little is known about 
pair production in non-constant magnetic fields \cite{Piazza,Ruf:2009zz}.
The clarification of potential, currently unknown phenomena associated with time-dependent magnetic fields is one of the required next steps if theoretical results on Schwinger pair production shall be put to the scrutiny of experiment. 

Among worldline \cite{Dunne:2005sx,Schneider:2014mla} and WKB-like formalism 
\cite{Strobel:2013vza}, 
the introduction of quantum kinetic theory \cite{Smolyansky:1997fc}
has helped to understand pair production in homogeneous, but time-dependent 
electric fields. (NB: For recent developments concerning quantum kinetic theory
see, {\it e.g.}, refs.\ \cite{Dabrowski:2014ica,Kohlfurst:2012rb,Hebenstreit:2014lra,Diss,Li:2014nua,Hebenstreit2015,Blinne2015}).
However, to accurately describe pair production in laser fields
one has to take into account spatial inhomogeneities 
\cite{Dunne:2005sx,Berenyi:2013eia,Hebenstreit:2011wk,Han:2010rg,Harvey:2012ie}
as well as magnetic fields \cite{Ruf:2009zz}.
In this letter, we will discuss the results of our exploratory study on the
influence of time-dependent, spatially inhomogeneous magnetic fields on the particle production rate using still a relatively simple model for the gauge potential.
To put these results into perspective, 
we will also compare the outcome of these calculations with a field configuration not fulfilling the homogeneous Maxwell equations. Our results are based upon the 
Dirac-Heisenberg-Wigner (DHW)
approach \cite{Vasak:1987um},
which was successfully employed for spatially inhomogeneous electric fields only
recently \cite{Berenyi:2013eia,Hebenstreit:2011wk}.

\paragraph{Formalism}

Throughout this article the convention $c = \hbar = m = 1$ will be used.
The theoretical approach employed here is based on the 
{fundament} laid by {refs}.~\cite{Vasak:1987um}.

{
The fundamental quantity in the DHW approach is the covariant Wigner operator
\begin{equation}
 \hat{\mathcal W}_{\alpha \beta} \left( r , p \right) = \frac{1}{2} \int d^4 s \ \mathrm{e}^{\mathrm{i} ps} \ \mathcal U \left(A,r,s \right) \ \mathcal C_{\alpha \beta} \left( r , s \right),
\end{equation}
where we have introduced the density operator
\begin{equation}
 \mathcal C_{\alpha \beta} \left( r , s \right) = \left[ \bar \psi_\beta \left( r - s/2 \right), \psi_\alpha \left( r + s/2 \right) \right]
\end{equation}
and the Wilson line factor
\begin{equation}
 \mathcal U \left(A,r,s \right) = \exp \left( \mathrm{ie} \int_{-1/2}^{1/2} d \psi \ A \left(r+ \psi s \right) \ s \right).
\end{equation}
The vector potential $A$ is given in mean-field approximation, 
$r$ and $s$ denote center-of-mass and relative coordinates, respectively. 
Taking the vacuum expectation value of the Wigner operator and projecting on equal time ({\it i.e.}, performing an integral $\int dp_0$) yields the single-time Wigner function $\mathcal W \left( \mathbf x , \mathbf p , t\right)$.
}

{
The simplest way to incorporate inhomogeneous magnetic fields is to investigate pair production in the $xz$-plane.
However, there are in total three different ways of defining
the basis matrices for a DHW calculation with only two spatial dimensions: one 
representation using 4-spinors and two representations using 2-spinors. Generally,
the 4-spinor formulation contains all information on the pair production process,
while the results from a 2-spinor formulation are spin-dependent (one describes
electrons with spin up and positrons with spin down \cite{deJesusAnguianoGalicia:2005ta}
and the other describes the spin-reversed particles).
}

{
To simplify the calculations we use a 2-spinor representation. Hence, we decompose the Wigner function into Dirac bilinears:
\begin{equation}
 \mathcal W \left( \mathbf x, \mathbf p, t \right) = \frac{1}{2} \left( \mathbbm 1 \ \mathbbm s + \gamma_{\mu} \mathbbm v^{\mu} \right).
\end{equation}
Following refs. \cite{Vasak:1987um} we are able to identify $\mathbbm s$ as mass density and $\mathbbm v^{\mu}$ as charge
and current densities.
}

We can reduce the corresponding equations of motions
for the Wigner {coefficients} $\mathbbm{s}$ and {$\mathbbm{v}^\mu = 
(\mathbbm{v}_0,\mathbbm{v}^1,\mathbbm{v}^3)$} to the form
(see, {\it e.g.}, {ref}.\ \cite{Diss}): 
\begin{alignat}{4}
  & D_t \mathbbm{v}_0     && +\mathbf{D} \cdot \mathbbm{v} && &&= 0, 
\label{eqn1_1} \\  
  & D_t \mathbbm{s}     && && -2 \left( \Pi_x \cdot \mathbbm{v}^3 -
\Pi_z \cdot \mathbbm{v}^1 \right) &&= 0,  \label{eqn1_2} \\  
  & D_t \mathbbm{v}^1    && +D_x \cdot \mathbbm{v}_0  && -2 \Pi_z
\cdot \mathbbm{s} && = -2 \mathbbm{v}^3,  \label{eqn1_3} \\  
  & D_t \mathbbm{v}^3     && +D_z \cdot \mathbbm{v}_0 && +2\Pi_x
\cdot \mathbbm{s} &&= 2 \mathbbm{v}^1,  \label{eqn1_4} 
\end{alignat}    
with the pseudo-differential operators
\begin{alignat}{8}
  & D_t && = \quad && \partial_{t} && + e && \int d\xi \, && \mathbf{E} \left(
\mathbf{x}+ \textrm {i} \xi \boldsymbol{\nabla}_p,t \right) && \cdot && \boldsymbol{\nabla}_p,
\label{eqn2_1}  \\
  & \mathbf{D} && = \quad && \boldsymbol{\nabla}_x && + e && \int d\xi \, && \mathbf{B}
\left( \mathbf{x}+\textrm {i} \xi \boldsymbol{\nabla}_p,t \right) && \times && \boldsymbol{\nabla}_p,
\label{eqn2_2}  \\
  & \boldsymbol{\Pi} && = \quad && \mathbf{p} && - \textrm {i} e && \int d\xi \,
\xi \, && \mathbf{B} \left( \mathbf{x}+\textrm {i} \xi \boldsymbol{\nabla}_p,t \right) &&
\times && \boldsymbol{\nabla}_p. \label{eqn2_3} 
\end{alignat}
The vacuum initial conditions are given by 
\begin{alignat}{3}
  \mathbbm{s}_{vac} \left(\boldsymbol{p} \right) = -\frac{2}{\sqrt{1 +
\boldsymbol{p}^2}}, \quad && 
  \mathbbm{v}_{vac}^{1,3} \left(\boldsymbol{p} \right) = -\frac{2
\boldsymbol{p}}{\sqrt{1 + \boldsymbol{p}^2}}. \label{eqn3}
\end{alignat}
For later use {we} explicitly subtract the vacuum terms by defining 
\begin{align}
 \mathbbm{w}^v = \mathbbm{w} - \mathbbm{w}_{vac},
\end{align}
{with $\mathbbm{w} = \mathbbm{v}_0 ,~\mathbbm{s} ,~\mathbbm{v}^1$ and
$\mathbbm{v}^3$, respectively \cite{Hebenstreit:2011pm}.}
The particle number density in momentum space is given by
\begin{align}
n \left( p_x, p_z \right) = \int dz \, \frac{\mathbbm{s}^v + p_x
\mathbbm{v}^{v,1} + p_z \mathbbm{v}^{v,3}}{\sqrt{1+\boldsymbol{p}^2}}.
 \label{eqn5}
\end{align}
When evaluated at asymptotic times, this quantity gives the particle momentum
spectrum. Subsequently,
the particle yield per unit volume element is obtained via $N = \int dp_x \, dp_z \, n \left( p_x ,p_z
\right)$.

{
In the following we will discuss pair production for one specific
2-spinor representation. The results for
particles with opposite spin can be obtained performing $p_z \to - p_z$.}

\paragraph{Solution strategies}

As momentum derivatives appear as arguments of $\mathbf{E}$ and $\mathbf{B}$
we Taylor-expand 
the pseudo-differential operators in \eqref{eqn2_1}-\eqref{eqn2_3} up to {fourth
order \cite{Diss}}. To increase numerical stability canonical momenta are used:
\begin{equation}
 \boldsymbol{q} = \boldsymbol{p} + e\boldsymbol{A} \left( \boldsymbol{x}, t \right) .
\end{equation}

In order to solve eqs.\ \eqref{eqn1_1}-\eqref{eqn1_4} numerically, 
spatial and momentum directions are equidistantly discretized, and {additionally we set
$\mathbbm{w}^v \left(\mathbf x_0 \right) = \mathbbm{w}^v \left(\mathbf x_{N_\mathbf{x}} \right) $
as well as $\mathbbm{w}^v \left(\mathbf p_0 \right) = \mathbbm{w}^v \left(\mathbf p_{N_\mathbf p} \right) $.} 
{We further demand Dirichlet boundary conditions
\begin{align}
 \mathbbm{w}^v \left( \mathbf x_{k_i}, \mathbf p_{k_j} \right) = 0 \quad \textrm{if} \quad
k_i = 0 \ \textrm{or} \ k_j = 0.
\end{align}
}
The derivatives are then calculated using pseudospectral methods in Fourier basis \cite{Boyd}.
The time integration was performed using a Dormand-Prince Runge-Kutta integrator of
order 8(5,3) \cite{NR}. 

\paragraph{Model for the fields}

For our studies of pair production in electromagnetic fields,
we choose a vector potential of the form
\begin{align}
 \boldsymbol{A} (z,t) &= \varepsilon \ \tau \left( \tanh \left(
\frac{t+\tau}{\tau} \right) - \tanh \left( \frac{t-\tau}{\tau} \right) \right)
\notag \\
  &\times \exp \left( -\frac{z^2}{2 \lambda^2} \right) \ \boldsymbol{e}_x. 
  \label{AA}
\end{align}
If not stated otherwise, the electric and magnetic field are derived from this expression. 
 {Note that 
the field configuration obeys $\mathcal G = \boldsymbol{E} \cdot \boldsymbol{B} = 0$. Moreover,} the homogeneous
Maxwell equations are
automatically fulfilled and additionally $\boldsymbol{\nabla} \cdot \boldsymbol{E} = 0$
holds.

The electric field is antisymmetric in time exhibiting a double peak structure
with $\varepsilon$
denoting the field strength. The field strength
of the magnetic field, however, is {suppressed relative to the electric field strength} by a term $\tau/
\lambda^2$,
where $\tau$ and $\lambda$ also determine the scale for temporal and spatial
variations, respectively.
Hence, for $\tau/ \lambda \ll 1$ the field energy is stored almost exclusively
in the
electric field. For $\tau/ \lambda \gtrsim 1$, however, the energy
stored in the magnetic field exceeds the energy fraction coming from the
electric part.

In ref.\ \cite{Dunne:2004nc} it was argued
that pair production is only
possible in regions where $\mathbf{E} \left( z,t \right)^2 - \mathbf{B} \left( z,t \right)^2 > 0$.
To analyze our results in view of this conjecture we therefore define an ``effective field amplitude''
\begin{equation}
 \tilde E \left( z,t \right)^2 = \mathbf{E} \left( z,t \right)^2- \mathbf{B} \left( z,t \right)^2
\end{equation}
and a ``modified effective field energy'' 
\begin{equation}
 \mathcal{E} \left( \mathbf{E}, \mathbf{B} \right) = \int \ dz \ dt \ \tilde E \left( z,t
\right)^2 \ \Theta \left(  \tilde E \left( z,t
\right)^2 \right),
 \label{EE}
\end{equation}
with the Heaviside function $\Theta \left(x \right)$.

\begin{figure}[h]
\centering
\includegraphics[width=0.49\textwidth]{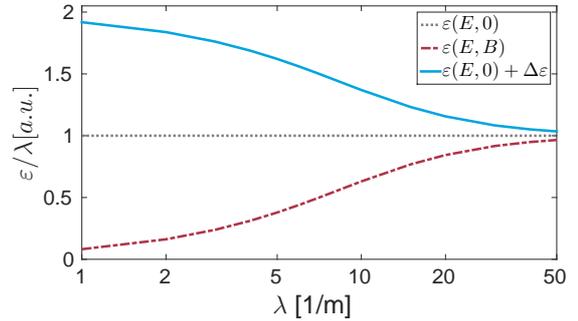}
\caption{Qualitative comparison of the electric field energy (dotted grey line)
and
the effective field energy using the proposed vector potential (dot-dashed red line) for
$\tau = 10/m$.
{In addition, we show for later comparison the difference in energy ($\Delta \varepsilon$) added to the electric field energy (blue line).} }
\label{fig:Inv}
\end{figure} 

This effective energy \eqref{EE} for different values of $\lambda$ is displayed in Fig.
\ref{fig:Inv}.
We find that the magnetic field can significantly reduce the effective field
strength. 
While the electric part linearly depends on $\lambda$, the calculation for
a combined electric and magnetic field shows a rapid drop off for $\tau/\lambda
\gtrsim 1$.

\paragraph{Particle distribution}

\begin{figure}[h]
        \includegraphics[width=0.49\textwidth]{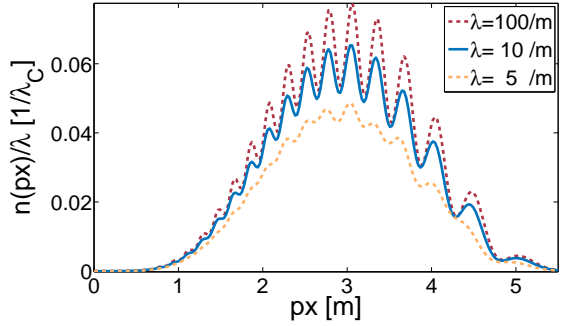}
        \caption{Reduced particle density $n(p_x)/ \lambda$ for various values of the spatial
inhomogeneity $\lambda$, a field strength of $e\varepsilon = 0.707 \, m^2$
and a pulse length $\tau = 5 /m$.
        For $\lambda \gg \tau$ the reduced particle density converges.}
        \label{fig:distr_x}
      \end{figure}

It is useful to define the reduced particle density $n(p_x)/ \lambda$ to scale out the 
trivial linear dependence on $\lambda$. As can be seen in Fig.~\ref{fig:distr_x},  
$n(p_x)/ \lambda$ displays a peaked structure  superimposed by an oscillating function. This
is characteristic for electric fields with peaks of the same absolute
value but opposite sign \cite{Hebenstreit:2009km}.
      
It should be pointed out that especially the peaks in the reduced particle
distribution $n(p_x)/ \lambda$ 
decrease with decreasing $\lambda$. 
A possible interpretation is that the presence of the magnetic field prevents
the particles,
created at the different field oscillations, to interfere. In case of
$\tau/\lambda \gtrsim 1$ particles 
created around the first electric field oscillation at $z \neq 0$ are
accelerated in $x$ and also $z$ direction. 
However, particles created at the second oscillation acquire a completely
different momentum signature and therefore both wave
packages become distinguishable.
Moreover, an analysis of our data indicates, that the particle distribution is
slowly shifted to lower momenta for small $\lambda$. The reason for this
phenomenon seems to be directly linked with the increase in the magnetic field strength.
For a configuration of the form \eqref{AA}, a decrease of the parameter $\lambda$
causes the region with maximal effective field amplitude to be shifted away from $t=0$.
Therefore this shift has a different origin compared to the previously discovered
particle self-bunching \cite{Hebenstreit:2011wk}. 

\begin{figure}[h]
	\includegraphics[width=0.49\textwidth]{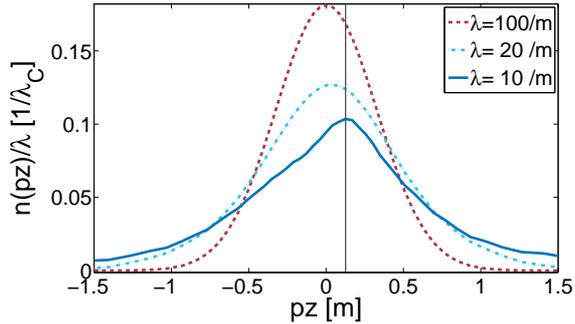} 
	\caption{Reduced particle density $n(p_z)/ \lambda$ for various values of the spatial
inhomogeneity $\lambda$.
	The particle density is symmetric for $\lambda \gg \tau$ only. The
vertical grey line is there to guide the eye: the peak of the particle density is
shifted to positive $p_z$. Parameters: $e\varepsilon = 0.707 \, m^2$
and $\tau = 5 /m$.}
	\label{fig:distr_z}
      \end{figure}

The reduced particle density $n(p_z)/ \lambda$, {\it cf.} Fig.~\ref{fig:distr_z}, 
does not show any
interference pattern. For $\lambda = 100/m$ the distribution in $p_z$ is symmetric around 
the origin, in agreement with homogeneous calculations. However, in case of $\tau/\lambda
\gtrsim 1$ the particle peak is shifted towards positive $p_z$. 

\newpage
As noted above,
using the second $2$-spinor basis, one obtains a  particle density mirrored at
$p_z=0$. 
Therefore, this result is an indicator for interactions
between the magnetic field and the electron spin. 
  
\paragraph{Particle yield}

The magnetic field is not independent of the electric field, because both stem
from the same vector potential \eqref{AA}. The effect of fixing $\mathbf{B} = 0$ and therefore
violating
the homogeneous Maxwell equation $\boldsymbol{\nabla} \times \boldsymbol{E} = -
\dot{\boldsymbol{B}}$
shows up in the particle density and subsequently in the particle yield.
In order to draw a general 
conclusion between effective field energy and particles created, we will focus on
the particle yield in the following.

      \begin{figure}[h]
	\includegraphics[width=0.49\textwidth]{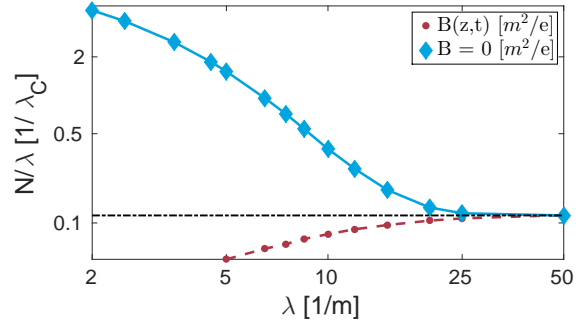}
	\caption{Double-log plot of the reduced particle yield as a function of the parameter $\lambda$. For $\lambda \gg \tau$ the reduced particle yield
	converges to the homogeneous result (dashed black line). In case of a sizable magnetic field the calculation for $B = 0$ (blue
line) leads to an overestimation compared to the correct result (dashed red line). Parameters: $\tau = 10/m$ and $e\varepsilon = 0.707 \, m^2$.}
	\label{fig:N1}
      \end{figure}

\begin{figure}[h]
	\includegraphics[width=0.49\textwidth]{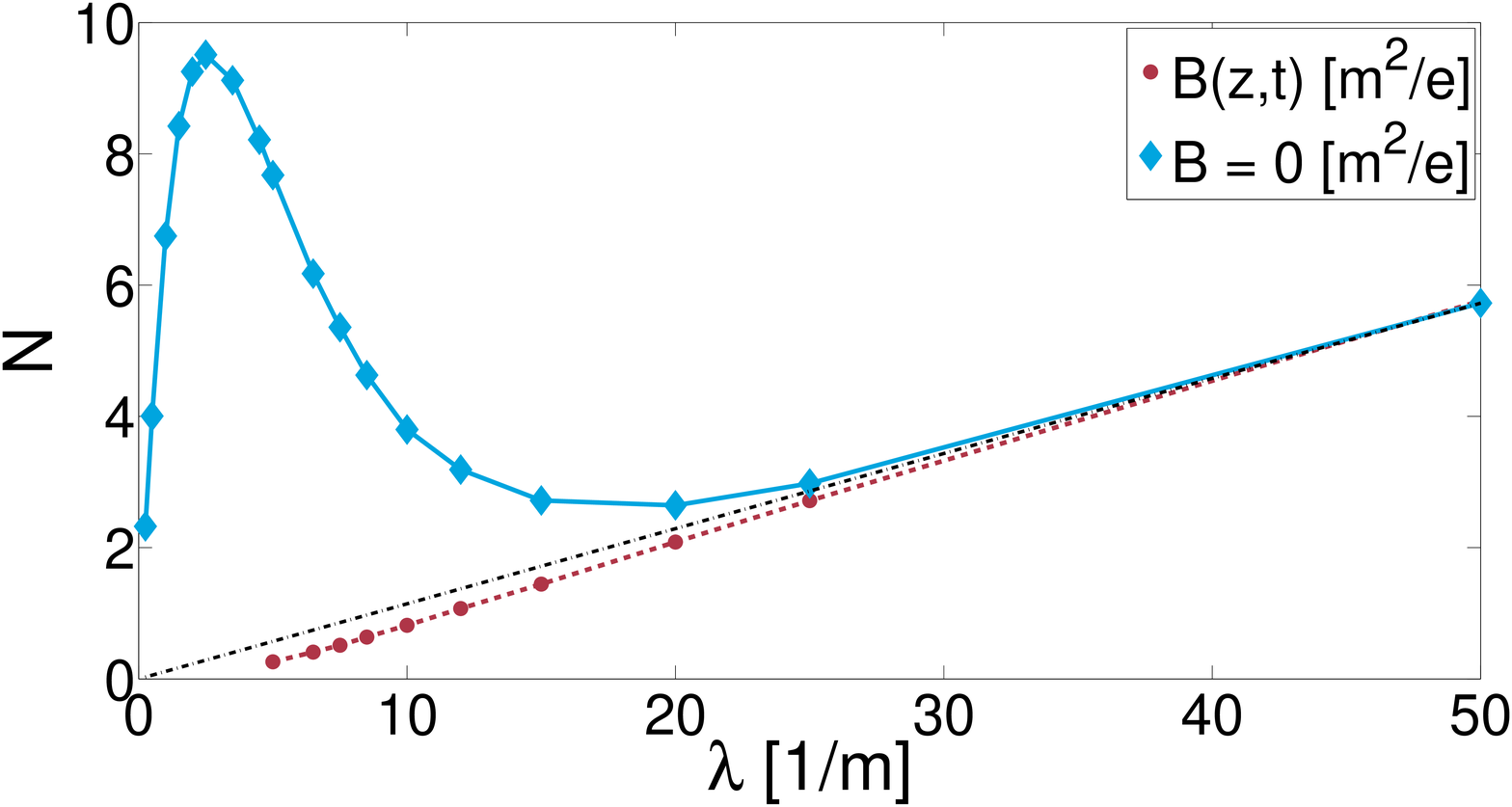}
	\caption{Particle yield drawn in a lin-lin plot for a configuration with $\tau = 10/m$ and $e\varepsilon = 0.707 \, m^2$ 
	(the same set of data as in Fig. \ref{fig:N1} is used). At $\lambda = 5/m$ the
particle yield obtained from a calculation
	with $B=0$ exceeds the correct result by a factor of $50$. }
	\label{fig:N2}
      \end{figure}

\newpage
The effective field energy without a magnetic field $\mathcal{E} \left(E,0
\right)$ is a linear function of $\lambda$, see Fig. \ref{fig:Inv}.
Hence, it is reasonable to introduce an approximation for the particle yield
\begin{align}
 \tilde{N} = \lambda \, N_{hom},
\end{align}
where $N_{hom}$ is the yield obtained from a calculation with a spatially homogeneous field.
The figures Fig. \ref{fig:N1} and Fig. \ref{fig:N2} show, that there is good agreement between the approximation
and the 
full solution for $\lambda \gg \tau$. Reasons are, that in this case
the electric field can be considered as quasi-homogeneous.
Furthermore, the magnetic field energy
is by orders of magnitude smaller than its electric counterpart and therefore
negligible.

The effect of spatial restrictions on the electric field has already been
investigated in Ref. \cite{Dunne:2005sx,Hebenstreit:2011wk}.
In our case, also the effect of a magnetic field growing in strength for
decreasing $\lambda$, has to be
taken into account. 
The corresponding computation of the effective field amplitude is depicted in Fig. \ref{fig:Inv} as $\mathcal{E}
\left(E, B \right)$. One observes a faster than linear decrease. This is in qualitative agreement with
the particle yield, as illustrated in Fig. \ref{fig:N1}.
We have to admit, however, that for calculations with $\lambda < 5/m$ the
results are not reliable
anymore due to a breakdown of the used Taylor expansion. {(NB: 
The calculation with $B=0$
does not display this numerical problem.)}

Eventually, the configuration $\boldsymbol{E} = - \dot{\boldsymbol{A}}$ and
$\boldsymbol{B} = 0$ is analyzed.
Contrary to the previous case, simply calculating the effective field energy of
the applied field, which would be
$\mathcal{E} \left( E,0 \right)$ in Fig. \ref{fig:Inv}, is not sufficient.
We have to consider, that the homogeneous Maxwell equations are not
fulfilled. Hence, we suggest to add the missing part of the effective field energy to the electric field
energy, {illustrated as $\mathcal{E} \left(E,0 \right)
+ \Delta \mathcal{E}$ in Fig. \ref{fig:Inv}.} 

In this way, the increase in the particle yield in Fig. \ref{fig:N1} and Fig. \ref{fig:N2} can be understood in terms
of the magnetic field. We assume, that a magnetic field hinders matter creation.
Fixing $B$ to zero and ignoring the term \mbox{$\boldsymbol{\nabla} \times \boldsymbol{A}$} in the
equations \eqref{eqn1_1}-\eqref{eqn1_4} therefore inevitably leads to an overestimation of the
effective field amplitude and consequently to an overestimation of the total particle number. 
(NB: Comparison of Fig. \ref{fig:Inv} with Fig. \ref{fig:N1} corroborates this argument.)

The sharp drop off on the left side of Fig. \ref{fig:N2} is connected to the
fact, that for $\lambda \to 0$
the energy stored in the background field is not sufficient anymore to overcome
the particle rest mass \cite{Hebenstreit:2011pm,Dunne:2005sx}.

\paragraph{Conclusions}

Based on the DHW formalism, Sauter-Schwinger electron-positron pair
production in time-dependent, spatially inhomogeneous
electric and magnetic fields has been investigated. For the
first time the equations of motion for an effectively $2+1$ dimensional
system has been solved numerically. 
We have focused on the influence of the magnetic field on
the pair production process for a special class of vector
potentials. We have found that for this kind of potentials
the magnetic field is of minor importance for a wide
range of parameter sets thereby validating studies which have been
performed so far. 
Additionally, there is perfect agreement in the results when
comparing to quantum kinetic theory in the limit of spatially homogeneous fields.
However, and as most important result presented here, we have verified
in a quantitative manner that
in the case of spatially strongly localized fields the results
can be explained assuming that pair production is
only possible in regions where the electric field exceeds the magnetic field.
In this parameter region the correct treatment of the magnetic field is
of utter importance.

\paragraph{Outlook}

In order to investigate pair production in background fields 
with more realistic length and time scales, improvements in the employed numerical 
methods will be necessary. {Such work is in progress, and it will allow to investigate more general electromagnetic fields. A possible extension would be, for example, the study of multi-photon pair production.
Therefore we are optimistic that the investigation 
presented here will soon serve as a basis for 
studies employing fields closer to experimentally feasible conditions.}


\paragraph{Acknowledgements}
We are grateful to Florian Hebenstreit and Daniel Ber\'{e}nyi for helpful discussions,
especially those about numerical methods used in this study. 
We thank Holger Gies and Alexander Blinne for many interesting discussions and a 
critical reading of this manuscript.\\
C. K.\ acknowledges funding by the Austrian Science Fund, FWF, through the 
Doctoral Program ``On Hadrons in Vacuum, Nuclei and Stars''(FWF DK W1203-N16) and
by BMBF under grant No.  05P15SJFAA (FAIR-APPA-SPARC).
We thank the research core area ``Modeling and Simulation'' for support.



\begin{thebibliography}{100} 
  
\bibitem{Sauter:1931zz} 
  F.~Sauter,
  Z.\ Phys.\  {\bf 69}, 742 (1931);
  W.~Heisenberg and H.~Euler,
  Z.\ Phys.\  {\bf 98}, 714 (1936);
  J.~S.~Schwinger,
  Phys.\ Rev.\  {\bf 82}, 664 (1951);
  
\bibitem{Burke:1997ew}
  D.~L.~Burke, R.~C.~Field, G.~Horton-Smith, T.~Kotseroglou, J.~E.~Spencer,
D.~Walz, S.~C.~Berridge and W.~M.~Bugg {\it et al.},
  Phys.\ Rev.\ Lett.\  {\bf 79} (1997) 1626;
  C. Bamber, S. J. Boege, T. Koffas, T. Kotseroglou, A. C. Melissinos, D. D. Meyerhofer {\it et al.},
  Phys. Rev. D 60, 092004 (1999);
  
\bibitem{eli}
  Proposal for a European Extreme Light Infrastructure (ELI),
  http://www.eli-laser.eu/;
  
\bibitem{Heinzl:2008an}
  T.~Heinzl and A.~Ilderton,
  Eur.\ Phys.\ J.\ D {\bf 55} (2009) 359
  [arXiv:0811.1960 [hep-ph]];
  
\bibitem{xfel}
  XFEL, http://www.xfel.eu/de/, 
  The HIBEF project, http://www.hzdr.de/db/Cms?pNid=427\&pOid=35325;
  
\bibitem{Ringwald:2001ib}
  A.~Ringwald,
  Phys.\ Lett.\ B {\bf 510} (2001) 107
  [hep-ph/0103185];
  
\bibitem{Marklund:2008gj}
  M.~Marklund and J.~Lundin,
  Eur.\ Phys.\ J.\ D {\bf 55} (2009) 319
  [arXiv:0812.3087 [hep-th]];
  O.~J.~Pike, F.~Mackenroth, E.~G.~Hill and S.~J.~Rose,
  Nature Photon.\  (2014);

\bibitem{Dunne:2004nc}
  G.~V.~Dunne,
  In *Shifman, M. (ed.) et al.: From fields to strings, vol. 1* 445-522
  [hep-th/0406216];
  

  
\bibitem{Alkofer:2001ik} 
  R.~Alkofer, M.~B.~Hecht, C.~D.~Roberts, S.~M.~Schmidt and D.~V.~Vinnik,
  Phys.\ Rev.\ Lett.\  {\bf 87}, 193902 (2001);
  
\bibitem{PhysRevLett.101.130404}
  R. Sch\"utzhold, H. Gies and G. Dunne,
  Phys.\ Rev.\ Lett.\  {\bf 101} (2008) 130404.
  
\bibitem{abc}
  C.~Kohlf\"urst, H.~Gies and R.~Alkofer,
  Phys.\ Rev.\ Lett.\  {\bf 112} (2014) 050402
  [arXiv:1310.7836 [hep-ph]];
  A.~Blinne and H.~Gies,
  Phys.\ Rev.\ D {\bf 89} (2014), 085001
  [arXiv:1311.1678 [hep-ph]];
    A.~Huet, S.~P.~Kim and C.~Schubert,
  Phys. Rev. D {\bf 90} (2014) 125033
  [arXiv:1411.3074 [hep-th]];
  
  I.~Akal, S.~Villalba-Ch{\'a}vez and C.~M\"uller,
  Phys. Rev. D {\bf 90} (2014) 113004
  [arXiv:1409.1806 [hep-ph]];
  A.~Otto, D.~Seipt, D.~Blaschke, B.~Kampfer and S.~A.~Smolyansky,
  Phys. Lett. B {\bf 740} (2015) 335
  [arXiv:1412.0890 [hep-ph]];    
  
\bibitem{Dunne:1997kw}
  D.~Cangemi, E.~D'Hoker and G.~V.~Dunne,
  Phys.\ Rev.\ D {\bf 52} (1995) 3163
  [hep-th/9506085];
  G.~V.~Dunne and T.~M.~Hall,
  Phys.\ Lett.\ B {\bf 419} (1998) 322
  [hep-th/9710062];
  S.~P.~Kim and D.~N.~Page,
  Phys.\ Rev.\ D {\bf 75} (2007) 045013
  [hep-th/0701047];
  R.~Ruffini, G.~Vereshchagin and S.~S.~Xue,
  Phys.\ Rept.\  {\bf 487} (2010) 1
  [arXiv:0910.0974 [astro-ph.HE]];
  M. Jiang, Q. Z. Lv, Y. Liu, R. Grobe and Q. Su,
  Phys. Rev. A {\bf 90} (2014) 032101;
  {A. Ilderton, G. Torgrimsson and J. W\aa{}rdh,
  Phys. Rev. D {\bf 92} (2015) 065001,
  [arXiv:1506.09186 [hep-th]];}
  
\bibitem{Nousch:2012xe}
  F.~Mackenroth and A.~Di Piazza,
  Phys.\ Rev.\ A {\bf 83} (2011) 032106
  [arXiv:1010.6251 [hep-ph]];
  T.~Nousch, D.~Seipt, B.~Kampfer and A.~I.~Titov,
  Phys.\ Lett.\ B {\bf 715} (2012) 246;
  
\bibitem{Diss}  
  C. Kohlf\"urst,
  PhD Thesis (2015)
  [arXiv:1512.06082 [hep-ph]];
  
  \bibitem{Kohlfurst:2013ura}
  C.~Kohlf\"urst, H.~Gies and R.~Alkofer,
  Phys.\ Rev.\ Lett.\  {\bf 112} (2014) 050402,
  [arXiv:1310.7836 [hep-ph]];
  
\bibitem{Ruf:2009zz}
  M.~Ruf, G.~R.~Mocken, C.~Muller, K.~Z.~Hatsagortsyan and C.~H.~Keitel,
  Phys.\ Rev.\ Lett.\  {\bf 102} (2009) 080402
  [arXiv:0810.4047 [physics.atom-ph]];
  
\bibitem{Hebenstreit:2011pm}
  F.~Hebenstreit,
  arXiv:1106.5965 [hep-ph];

\bibitem{Cohen:2008wz}
  T.~D.~Cohen and D.~A.~McGady,
  Phys.\ Rev.\ D {\bf 78} (2008) 036008
  [arXiv:0807.1117 [hep-ph]];
  F. Gelis and N. Tanji,
  arXiv: 1510.05451 [hep-ph];

\bibitem{Orthaber:2011cm}
  M.~Orthaber, F.~Hebenstreit and R.~Alkofer,
  Phys.\ Lett.\ B {\bf 698} (2011) 80
  doi:10.1016/j.physletb.2011.02.053
  [arXiv:1102.2182 [hep-ph]].

\bibitem{Linder:2015vta}
  M.~F.~Linder, C.~Schneider, J.~Sicking, N.~Szpak and R.~Schützhold,
  Phys.\ Rev.\ D {\bf 92} (2015) 085009
  doi:10.1103/PhysRevD.92.085009
  [arXiv:1505.05685 [hep-th]].

\bibitem{Panferov:2015yda}
  A.~D.~Panferov, S.~A.~Smolyansky, A.~Otto, B.~Kaempfer, D.~Blaschke and L.~Juchnowski,
  arXiv:1509.02901 [quant-ph].


\bibitem{Piazza}
  A. Di Piazza and G. Calucci,
  Astroparticle Physics {\bf 24} (2006) 520;
 
\bibitem{Dunne:2005sx}
  G.~V.~Dunne and C.~Schubert,
  Phys.\ Rev.\ D {\bf 72} (2005) 105004
  [hep-th/0507174];
  H.~Gies and K.~Klingmuller,
  Phys.\ Rev.\ D {\bf 72} (2005) 065001
  [hep-ph/0505099];
  
\bibitem{Schneider:2014mla}
  C.~Schneider and R.~Sch\"utzhold,
  arXiv:1407.3584 [hep-th];
  
\bibitem{Strobel:2013vza}
  C.~K.~Dumlu and G.~V.~Dunne,
  Phys.\ Rev.\ Lett.\  {\bf 104} (2010) 250402
  [arXiv:1004.2509 [hep-th]];
  H.~Kleinert and S.~S.~Xue,
  Annals Phys.\  {\bf 333} (2013) 104
  [arXiv:1207.0401 [physics.plasm-ph]];
  E.~Strobel and S.~S.~Xue,
  Nucl.\ Phys.\ B {\bf 886} (2014) 1153
  [arXiv:1312.3261 [hep-th]];
  
\bibitem{Smolyansky:1997fc} 
  S.~A.~Smolyansky, G.~Ropke, S.~M.~Schmidt, D.~Blaschke, V.~D.~Toneev and
A.~V.~Prozorkevich,
  hep-ph/9712377;
  Y.~Kluger, E.~Mottola and J.~M.~Eisenberg,
  Phys.\ Rev.\ D {\bf 58}, 125015 (1998);
  S.~M.~Schmidt, D.~Blaschke, G.~Ropke, S.~A.~Smolyansky, A.~V.~Prozorkevich and
V.~D.~Toneev,
  Int.\ J.\ Mod.\ Phys.\ E {\bf 7}, 709 (1998);
  J.~C.~R.~Bloch, V.~A.~Mizerny, A.~V.~Prozorkevich, C.~D.~Roberts,
S.~M.~Schmidt, S.~A.~Smolyansky and D.~V.~Vinnik,
  Phys.\ Rev.\ D {\bf 60} (1999) 116011
  [nucl-th/9907027];
  
\bibitem{Dabrowski:2014ica}
  R.~Dabrowski and G.~V.~Dunne,
  Phys.\ Rev.\ D {\bf 90} (2014) 025021
  [arXiv:1405.0302 [hep-th]];
  
\bibitem{Kohlfurst:2012rb}
  C.~Kohlfurst, M.~Mitter, G.~von Winckel, F.~Hebenstreit and R.~Alkofer,
  Phys.\ Rev.\ D {\bf 88} (2013) 045028
  [arXiv:1212.1385 [hep-ph]];
  
\bibitem{Hebenstreit:2014lra}
  F.~Hebenstreit and F.~Fillion-Gourdeau,
  Phys.\ Lett.\ B {\bf 739} (2014) 189
  [arXiv:1409.7943 [hep-ph]];
  
\bibitem{Hebenstreit2015}
  F. Hebenstreit,
  arXiv:1509.08693 [hep-ph];
  
\bibitem{Blinne2015}
  A. Blinne and E. Strobel,
  arXiv:1510.02712 [hep-ph];

\bibitem{Li:2014nua}
  Z.~L.~Li, D.~Lu, B.~F.~Shen, L.~B.~Fu, J.~Liu and B.~S.~Xie,
  arXiv:1410.6284 [hep-ph];
  
\bibitem{Berenyi:2013eia}
  D.~Ber\'{e}nyi, S.~Varr\'{o}, V.~V.~Skokov and P.~L\'{e}vai,
  Phys.\ Lett.\ B {\bf 749} (2015) 210
  [arXiv:1401.0039 [hep-ph]];
  
\bibitem{Hebenstreit:2011wk}
  F.~Hebenstreit, R.~Alkofer and H.~Gies,
  Phys.\ Rev.\ Lett.\  {\bf 107} (2011) 180403
  [arXiv:1106.6175 [hep-ph]];
 
\bibitem{Han:2010rg}
  H.~Kleinert, R.~Ruffini and S.~S.~Xue,
  Phys.\ Rev.\ D {\bf 78} (2008) 025011
  [arXiv:0807.0909 [hep-th]];
  W.~B.~Han, R.~Ruffini and S.~S.~Xue,
  Phys.\ Lett.\ B {\bf 691} (2010) 99
  [arXiv:1004.0309 [hep-ph]];
  F.~Hebenstreit, J.~Berges and D.~Gelfand,
  Phys.\ Rev.\ D {\bf 87} (2013) 105006
  [arXiv:1302.5537 [hep-ph]];
  
\bibitem{Harvey:2012ie}
  C.~Harvey, T.~Heinzl, A.~Ilderton and M.~Marklund,
  Phys.\ Rev.\ Lett.\  {\bf 109} (2012) 100402
  [arXiv:1203.6077 [hep-ph]];
  
\bibitem{Vasak:1987um}
  D.~Vasak, M.~Gyulassy and H.~T.~Elze,
  Annals Phys.\  {\bf 173} (1987) 462;
  I. Bialynicki-Birula, P. G\'ornicki and J. Rafelski, 
  Phys. Rev. D 44 (1991);
  P. Zhuang, U. Heinz, 
  Ann.Phys.245:311-338,1996
  [arXiv:nucl-th/9502034];   
  
\bibitem{deJesusAnguianoGalicia:2005ta}
  M.~de Jesus Anguiano Galicia and A.~Bashir,
  Few Body Syst.\  {\bf 37} (2005) 71
  [hep-ph/0502089];
  
\bibitem{Boyd}
  J. P. Boyd, 
  Dover Books on Mathematics (2001),
  ISBN : 9780486411835;
  
\bibitem{NR}
  W. H. Press, S. A. Teukolsky, W. T. Vetterling and B. P. Flannery
  Cambridge University Press
  ISBN-13: 978-0521880688;

\bibitem{Hebenstreit:2009km}
  F.~Hebenstreit, R.~Alkofer, G.~V.~Dunne and H.~Gies,
  Phys.\ Rev.\ Lett.\  {\bf 102} (2009) 150404
  [arXiv:0901.2631 [hep-ph]];
  E.~Akkermans and G.~V.~Dunne,
  Phys.\ Rev.\ Lett.\  {\bf 108} (2012) 030401
  [arXiv:1109.3489 [hep-th]];
  {T. Heinzl, A. Ilderton and M. Marklund,
  Phys. Lett. B {\bf 692} (2010) 250,
  [arXiv:1002.4018 [hep-ph]];}
  
\end{thebibliography}

\end{document}